# Efficient broadband infrared absorbers based on core-shell nanostructures

**HENRIK A. PARSAMYAN,[1] KHACHATUR V. NERKARARYAN [1,\*], AND SERGEY I. BOZHEVOLNYI[2]**

[1] *Department of Radiophysics, Yerevan State University, A. Manoogian 1, Yerevan, 0025, Armenia*
[2] *Centre for Nano Optics, University of Southern Denmark, Campusvej 55, DK-5230 Odense M, Denmark*
*\*Corresponding author: knerkar@ysu.am*

**Abstract:** We suggest to exploit dielectric-metal core-shell nanostructures for efficient resonant and yet broadband absorption of infrared radiation with deeply subwavelength configurations. Realizing that nanostructures would efficiently absorb radiation only when their dielectric properties match those of the environment and making use of the effective medium approach, we develop the design strategy using core-shell nanostructures with very thin shells made of poor metals, i.e., metals having real and imaginary parts of their dielectric permittivities of the same order of magnitude. Analyzing in detail spherical and cylindrical core-shell nanostructures, we demonstrate that the resonant infrared absorption can be not only very efficient, i.e., with the absorption cross sections exceeding geometrical ones, but also broadband with the spectral width being of the order of the resonant wavelength. We obtain simple analytical expressions for the absorption resonances in spherical and cylindrical configurations that allow one to quickly identify the configuration parameters ensuring strong infrared absorption in a given spectral range. Relations to effective medium parameters obtained by the internal homogenization are established and discussed. We believe that our results can be used as practical guidelines for realization of efficient broadband infrared absorbers of subwavelength sizes desirable in diverse applications.

## 1. Introduction

Electromagnetic wave absorbers are characterized by the efficient radiation absorption at operating wavelengths, with the electromagnetic energy being transformed into ohmic heat or other forms of energy, so that no sizeable transmission or reflection is produced as a result of wave interaction with an absorber. Recent advances in plasmonics and metamaterials research together with rapid progress in nanotechnology, especially in nanofabrication techniques, resulted in the development of novel configurations of electromagnetic absorbers. The main operation principle of these absorbers is based on the phenomenon of localized plasmon resonance (LPR) often complemented with a metamaterial concept to achieve much smaller absorber volumes, sufficient performance, and design flexibility based on geometry rather than the materials used [1–4]. Absorbers in various parts of electromagnetic spectrum, from visible to microwave [5–8], have been widely investigated for a wide range of applications, including stealth technology, bolometers and thermal emitters [9–11]. Many characteristic absorbers, exhibiting, for example, polarization-insensitive, wide-angle, multi-band and broadband absorption have been realized using different configurations [12–14]. Moreover, multi-band [15] or broadband [16] absorbers featuring polarization-insensitive and wide-angle absorption simultaneously have also been demonstrated experimentally.

Although numerous metamaterial structures have been suggested as perfect infrared absorbers, due to the resonant nature of such structures the absorption efficiency decreases drastically away from the resonant wavelength, i.e., when the LPR condition is not met. Also the electromagnetic response of a metamaterial is defined by the collective response of all

"meta-atoms", so that a large array of carefully designed subwavelength "meta-atoms" is required. The resulting configuration may have complex structure and morphology being composed of various material layers to obtain desired absorptive properties [7,12,17]. Furthermore, efficient absorption of infrared radiation with metal-dielectric heterogeneous configurations becomes progressively difficult to achieve for longer wavelengths due to rapid increases in absolute values of real and imaginary parts of the dielectric permittivity of metals. One would therefore like to simplify the absorber configuration while keeping broadband, polarization and angle independent properties.

As far as we know, there are no universal approaches to design an absorber that would operate efficiently in a wide wavelength range. The problem is that, at low frequencies (long wavelengths), due to drastic increases of magnitudes of both real and imaginary parts of the dielectric permittivity of metals, unique properties of localized surface plasmons practically disappear along with well-defined LPRs. As a result, radiation absorption by individual small metal particles rapidly decreases in the infrared [18], while coherent scattering processes caused by the structural periodicity become also less efficient. Analysis of the absorption by small particles shows that efficient absorption occurs only when the dielectric constants of a particle and its environment are of the same order. To deal with this problem, one can either use materials with appropriate dielectric constants, or exploit composite materials for controlling effective dielectric constants of heterogeneous configurations. The second approach is more flexible and promising from the viewpoint of practical realization. In fact, various core-shell structures have already become objects of numerous studies exploiting unique design possibilities opened by combining dielectric properties of the shell and core, which are not accessible for those based on single-component materials or alloys [17,19-21].

Here, using the electrostatic approximation complemented with finite-element simulations we analyze in detail spherical and cylindrical core-shell nanostructures and demonstrate that the resonant infrared absorption can be not only very efficient, i.e., with the absorption cross sections exceeding geometrical ones, but also broadband with the spectral width being of the order of the resonant wavelength. We find that core-shell nanostructures with very thin shells made of poor metals, i.e., metals having real and imaginary parts of their dielectric permittivities of the same order of magnitude, can also be used for efficient broadband absorption of infrared radiation, and obtain simple analytical expressions for the absorption resonances in spherical and cylindrical configurations. Relations to effective medium parameters obtained by the internal homogenization are established and discussed.

## 2. Theory

In this section, we analyze spherical and cylindrical core-shell nanostructures within the framework of electric-dipole electrostatic approximation, considering thin metal shells with very large dielectric permittivities found typically for long (infrared) wavelengths.

### 2.1 Spherical configuration

Considering the spherical core-shell configuration, let $\varepsilon_1$, $\varepsilon_2$ and $\varepsilon_3$ be the dielectric permittivities of the core, shell and embedding medium, respectively. The core radius is $R_1$ and the shell thickness is $h = R_2 - R_1$, where $R_2$ is the total radius of the system. Within the limits of electrostatic approximation (i.e., for $R_2 \ll \lambda$, where $\lambda$ is the incident light wavelength in the surrounding free space) the absorption cross section (ACS) of the spherical structure is defined by the following expression [18,21]:

$$\sigma_{abs}^{sph} = \frac{8\pi^2}{\lambda} R_2^3 \Im m\left(X_{sph}\right) \qquad (1)$$

where

$$X_{\text{sph}} = \frac{(1+2\eta\mu)\varepsilon_2 - (1-\eta\mu)\varepsilon_3}{(1+2\eta\mu)\varepsilon_2 + (1-\eta\mu)\varepsilon_3}, \quad \eta = \frac{R_1^3}{R_2^3}, \quad \mu = \frac{\varepsilon_1 - \varepsilon_2}{\varepsilon_1 + 2\varepsilon_2} \tag{2}$$

Hereinafter we assume that the cores and surrounding media of the investigated systems are dielectrics and the shells are metals with the frequency-dependent dielectric permittivity $\varepsilon_2(\omega) = \varepsilon_{2r}(\omega) + i\varepsilon_{2i}(\omega)$, where $\varepsilon_{2r} < 0$.

In the investigated case, when $|\varepsilon_2| \gg \varepsilon_1, \varepsilon_3$, the imaginary part of $X_{\text{sph}}$ from Eq. (2), defining the absorption properties of the system, has the following form:

$$\Im m(X_{\text{sph}}) = \frac{18(1-\eta)\varepsilon_3\varepsilon_{2i}}{\left[3(\varepsilon_1 + 2\varepsilon_3) + 2(1-\eta)\varepsilon_{2r}\right]^2 + 4(1-\eta)^2 \varepsilon_{2i}^2}. \tag{3}$$

It is seen that, for given material properties and light wavelength, the absorption reaches its maximum when the geometrical parameters satisfy the following condition:

$$(1-\eta) = \frac{3(\varepsilon_1 + 2\varepsilon_3)}{2|\varepsilon_2|} \ll 1, \tag{4}$$

and the maximum value is as follows:

$$\{\Im m(X_{\text{sph}})\}_{\max} = \frac{3\varepsilon_3\varepsilon_{2i}}{2(\varepsilon_1 + 2\varepsilon_3)(|\varepsilon_2| + \varepsilon_{2r})}. \tag{5}$$

### 2.2 Cylindrical configuration

Let us now discuss the case of a dielectric cylinder covered with a metal layer. We assume that the electric field of the incident wave is perpendicular to the axis of the cylinder, i.e., the *p*-polarized incidence is considered. All notations and conditions introduced for the spherical configuration described above will also be used when analyzing the core-shell cylindrical configuration. Additionally, we assume that $R_2 \ll L \ll \lambda$, where $L$ is the cylinder length.

Within the electrostatic approximation the ACS of a core-shell cylinder is as follows:

$$\sigma_{\text{abs}}^{\text{cyl}} = \frac{8\pi^2 L}{\lambda} R_2^2 \Im m(X_{\text{cyl}}) \tag{6}$$

where

$$X_{\text{cyl}} = \frac{\varepsilon_3 - \varepsilon_2 + \zeta\gamma(\varepsilon_3 + \varepsilon_2)}{\varepsilon_3 + \varepsilon_2 + \zeta\gamma(\varepsilon_3 - \varepsilon_2)}, \quad \zeta = \frac{R_1^2}{R_2^2}, \quad \gamma = \frac{\varepsilon_1 - \varepsilon_2}{\varepsilon_2 + \varepsilon_1} \tag{7}$$

Again, taking into account the condition $|\varepsilon_2| \gg \varepsilon_1, \varepsilon_3$, the equation similar to Eq. (3) can be derived for the core-shell cylinders:

$$\Im m(X_{\text{cyl}}) = \frac{4(1-\zeta)\varepsilon_3\varepsilon_{2i}}{\left[2(\varepsilon_1 + \varepsilon_3) + (1-\zeta)\varepsilon_{2r}\right]^2 + (1-\zeta)^2 \varepsilon_{2i}^2}. \tag{8}$$

Continuing to use the same line of reasoning as above, one arrives at the following condition for the absorption to reach its maximum:

$$(1-\eta) = \frac{2(\varepsilon_1 + \varepsilon_3)}{|\varepsilon_2|} \ll 1, \quad (9)$$

with the corresponding maximum value:

$$\{\Im m(X_{cyl})\}_{max} = \frac{\varepsilon_3 \varepsilon_{2i}}{(\varepsilon_1 + 2\varepsilon_3)(|\varepsilon_2| + \varepsilon_{2r})}. \quad (10)$$

For convenience, we introduce the absorption and scattering efficiency factors [22]:

$$Q_{abs} = \frac{\sigma_{abs}}{S} \text{ and } Q_{scat} = \frac{\sigma_{scat}}{S} \quad (11)$$

where $S$ is the corresponding geometrical cross section area, which is $S_{sph} = \pi R_2^2$, for a sphere, and $S_{cyl} = 2R_2 L$, for a cylinder.

Let us consider the absorption of infrared radiation by core-shell configurations made of poor (such as Ti, Cr and Ni) metals, whose real and the imaginary parts of dielectric permittivities are of same order of magnitude, so that the following condition is fulfilled:

$$\varepsilon_{2i} \sim |\varepsilon_{2r}| \gg \varepsilon_1 + 2\varepsilon_3. \quad (12)$$

It is seen that, in this case and under the conditions of maximum absorption [cf. Eq. (4) and Eq. (9)], the absorption efficiencies even for very large permittivities can reach the unity and even greater values for subwavelength configurations, because $\text{Im}(X_{sph}) \sim \text{Im}(X_{cyl}) \sim 1$ [cf. Eq. (3) and Eq. (8)] resulting in $Q_{abs}^{sph} \sim 8\pi R_2/\lambda$ and $Q_{abs}^{cyl} \sim 4\pi^2 R_2/\lambda$. It should also be noted that, for both configurations, the absorption scales as $|\varepsilon_{2r}|/\varepsilon_{2i}$ and can benefit from relatively small imaginary parts of metal permittivity.

## 3. Numerical analysis

Numerical simulations are conducted using the finite-element method (FEM) for calculating the absorption and scattering efficiencies of core-shell spherical and cylindrical nanostructures with thin metal shells. First, we consider the theoretical and simulated absorption efficiency spectra of 500-nm-radius glass configurations covered with different titanium (Ti) [23] shell thicknesses surrounded by air [see Fig. (1)].

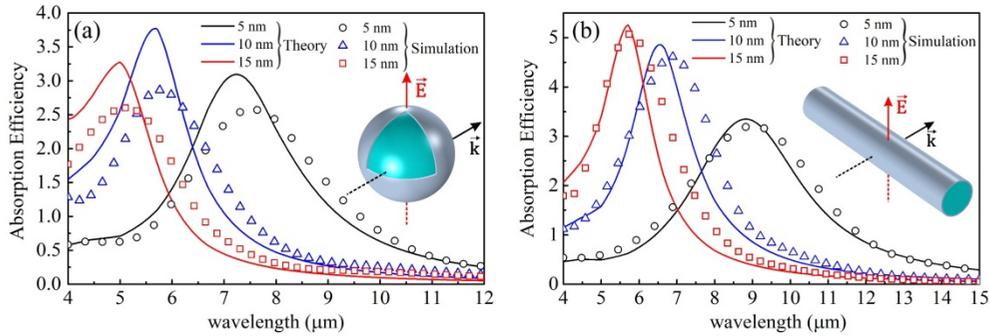

Fig. 1. Theoretical (solid lines) and simulated (dots) absorption efficiency spectra of the (a) spherical and (b) cylindrical glass-core and Ti-shell structures with the shell thicknesses of 5 nm (black), 10 nm (blue) and 15 nm (red). The core radius of both configurations is 500 nm and the glass permittivity is $\varepsilon_1 = 2.25$. The surrounding medium is air ($\varepsilon_3 = 1$). The insets show the considered configurations and incident plane wave characteristics, with the cylinder being oriented perpendicular to both electric field and propagation directions.

It is seen see that the theoretical and simulated absorption efficiency spectra for both configurations agree reasonably well, with their correspondence becoming better for longer wavelengths because of the electrostatic approximation becoming better justified. Both configurations, when decreasing the shell thickness, exhibit the same trends of the absorption maximum shifting to longer wavelengths and the bandwidth becoming wider. The first trend is related to the Ti-permittivity magnitude increasing for longer wavelengths [cf. Eq. (4) and Eq. (9)], while the second trend is due to improving (for longer wavelengths) the balance between the Ti-permittivity real and imaginary parts: $\varepsilon_2 = -213.85 + 89.2i$ at 7.6 $\mu$m and $\varepsilon_2 = -334.76 + 144.16i$ at 9.5 $\mu$m [23]. Note that, for both configurations, the absorption cross section exceeds the geometrical one in wide ranges of wavelengths, so that the absorption bandwidths are of the same order of magnitude as the resonance wavelength. For instance, for the shell thickness $h$ = 5 nm, Ti-coated spheres and cylinders feature the absorption efficiencies exceeding unity within the wavelength range of 6 to 9.5 $\mu$m [Fig. 1(a)] and 6.5 to 12 $\mu$m [Fig. 1(b)], respectively.

The absorption and scattering spectra calculated for various Ti-shell thicknesses (from 3 to 30 nm) are compared in Fig. 2, revealing that the resonant absorption becomes dominant for longer wavelengths with the (resonant) scattering becoming negligibly small for shell thicknesses below 10 nm [Fig. 2(b,d)]. As far as the possibility of infrared absorption is concerned, satisfying the condition of strong absorption [cf. Eq. (4) and Eq. (9)] for longer wavelengths requires progressively thinner shells [Fig. 2(a,c)], a circumstance that would eventually result in the limit on practical realization of strong absorption (and near-zero scattering) for long wavelengths (>10 $\mu$m).

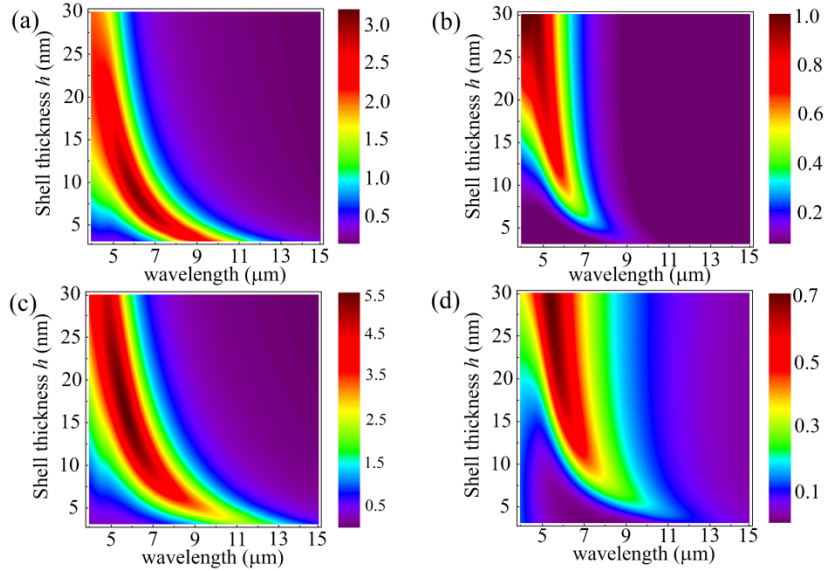

Fig. 2. Contour plots of the absorption and scattering efficiency spectra for spherical (a, b) and cylindrical (c, d) Ti-shell structures as a function of the shell thickness $h$. The radius of cores for both structures is 500 nm. All else is as in Fig. 1.

Finally, we would like to emphasize that the considered configurations are characterized by high degrees of symmetry, ensuring thereby that the absorption is completely (spherical) or partially (cylinder) insensitive to the angle and polarization of incident radiation. Unlike the case of traditional metamaterial absorbers in the infrared, there is no need for an orderly arrangement of these elements, an important feature that opens the prospect for practical use of these structures in random surface arrangements.

## 4. Effective medium approach

Analysis of the absorption by small homogeneous particles (characterized by the dielectric permittivity $\varepsilon$) shows that the absorption efficiencies rapidly decrease for very large permittivities because $\text{Im}(X_{\text{sph}})$ and $\text{Im}(X_{\text{cyl}})$ scale as $\text{Im}(\varepsilon)/|\varepsilon|^2$, so that the efficient absorption by a nanoparticle in air would occur only when the particle permittivity would be close to unity [18]. In fact, the underlying physical reason for the investigated structures to feature strong infrared absorption is in the realization of effective material properties close to unity. In order to reveal the physical meaning for the absorptive and scattering properties of considered core-shell configurations, let us substitute a core-shell sphere characterized by the dielectric constants $\varepsilon_1$ and $\varepsilon_2$ representing the core and shell, respectively, with a homogeneous sphere having the effective complex dielectric permittivity of $\varepsilon_{\text{eff}}^{\text{sph}}$ and the radius equal to the total radius of the core-shell sphere. Since the considered core-shell spheres are assumed to be of deeply subwavelength sizes, the electrostatic approximation can be applied to derive the effective medium permittivity. The effective permittivity can thereby be obtained by means of the internal homogenization based on Maxwell-Garnett mixing theory as follows [24]:

$$\varepsilon_{\text{eff}}^{\text{sph}} = \varepsilon_2 \frac{(\varepsilon_1 + 2\varepsilon_2) + 2\eta(\varepsilon_1 - \varepsilon_2)}{(\varepsilon_1 + 2\varepsilon_2) - \eta(\varepsilon_1 - \varepsilon_2)} \qquad (13)$$

where $\eta$ is defined in Eq. (2), and represents the volume fraction of a spherical core. The effective dielectric permittivity of a nano-cylinder, whose absorption and scattering properties would be identical with those of a cylindrical core-shell nanostructure can be established following the same route:

$$\varepsilon_{\text{eff}}^{\text{cyl}} = \varepsilon_2 \frac{(\varepsilon_1 + \varepsilon_2) + \zeta(\varepsilon_1 - \varepsilon_2)}{(\varepsilon_1 + \varepsilon_2) - \zeta(\varepsilon_1 - \varepsilon_2)}. \qquad (14)$$

where $\zeta$ is defined in Eq. (7), and represents the volume fraction of a cylindrical core.

The effective medium approximation was conducted for both spherical and cylindrical core-shell configurations considering the cores with the dielectric permittivity of $\varepsilon_1 = 2.25$ and radius $R_1 = 500$ nm covered by a 5-nm-thin Ti layer ($\varepsilon_2 = \varepsilon_{\text{Ti}}$), whose real and imaginary parts of dielectric permittivity [23] are shown in Fig. 3(a). The effective dielectric permittivities of equivalent sphere and cylinder calculated according to Eqs. (13) and (14) are displayed in Fig. 3(b). Solid lines stand for the real, while the dashed ones-for the imaginary parts of effective dielectric constants of the spherical (blue) and cylindrical (red) core-shells. The corresponding calculations demonstrate that the absorption and scattering efficiency spectra obtained for the core-shell configurations and effective-medium nanoparticles are in rather good agreement [cf. Fig. 3(d) and (e)], confirming thereby the validity of the effective medium approaches based on Eqs. (13) and (14). Importantly, it is immediately revealed that, as expected, the presence of dielectric spherical and cylindrical cores results in appropriately reducing both imaginary and absolute of the real parts of the effective dielectric permittivities compared to those of homogeneous Ti [cf. Fig. 3(a) and (b)]. For instance, at the strongest absorption by the spherical configuration is achieved at $\lambda = 7.6~\mu$m, at which the Ti dielectric permittivity is $\varepsilon_{\text{Ti}} = -213.85 + 89.2i$, while that of the effective medium $\varepsilon_{\text{eff, sph}} = -2.01 + 1.77i$. Similarly, at the absorption resonance wavelength of the cylindrical configuration $\lambda = 9.5~\mu$m, the Ti and effective medium dielectric permittivities are $\varepsilon_{\text{Ti}} = -334.76 + 144.16i$ and $\varepsilon_{\text{eff, cyl}} = -1.09 + 1.441i$, respectively.

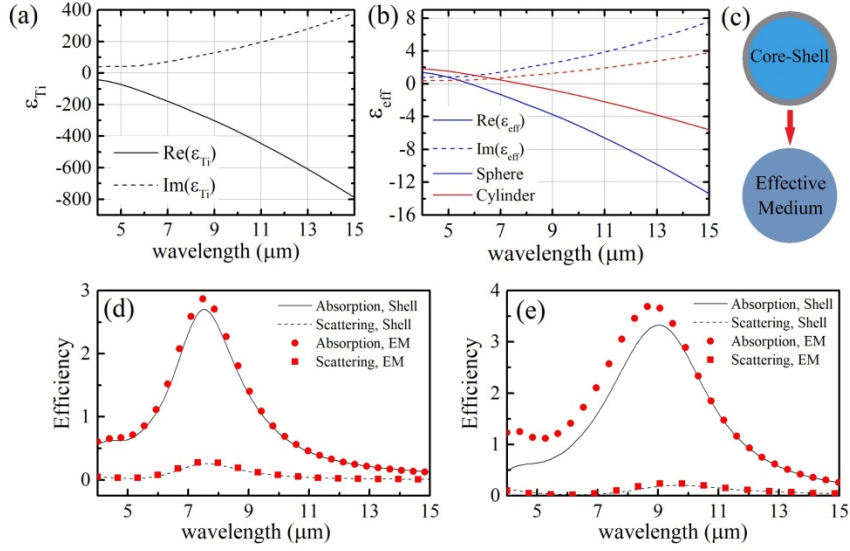

Fig. 3. (a) The real and the imaginary parts of dielectric permittivity of Ti shells [23]. (b) The real (solid) and imaginary (dashed) parts of effective dielectric permittivities of equivalent sphere (blue) and cylinder (red) calculated for the core-shell configurations with the core radius $R_1 = 500$ nm and dielectric constant $\varepsilon_1 = 2.25$ and 5-nm-thin Ti shell. (c) Schematic of the core-shell to effective-medium conversion. (d,e) The absorption (solid lines, circles) and scattering (dashed lines, squares) efficiency spectra of (d) spherical and (e) cylindrical core-shell and effective-medium (with the radius $R_{eff} = 505$ nm) configurations. Solid and dashed lines correspond to the core-shell configurations, while symbols (circles and squares) represent the effective-medium equivalents.

The resonant values of effective permittivities mentioned above for a homogeneous subwavelength sphere and cylinder are very close to those expected for the corresponding LSP resonances of homogeneous nanoparticles in air, viz., $Re(\varepsilon_{sph}) = -2$ and $Re(\varepsilon_{cyl}) = -1$, but noticeably different. This difference comes from the fact that the real and imaginary parts of the corresponding effective permittivities [cf. Eqs. (13) and (14)] are not independent quantities. Thus, within the approximation used that the metal permittivity is much larger than that of dielectrics, one obtains from Eq. (13) and the convention $\varepsilon_{eff} = \varepsilon_{effr} + i\varepsilon_{effi}$ the following relation between the real and imaginary parts of the sphere effective permittivity: $\varepsilon_{effr} = \varepsilon_1 + (\varepsilon_{2r}/\varepsilon_{2i})\varepsilon_{effi}$. Applying this relationship and searching for the maximum of the imaginary part of a sphere (electrostatic) polarizability, $Im[(\varepsilon_{eff} - \varepsilon_3)/(\varepsilon_{eff} + 2\varepsilon_3)]$, results immediately in recovering the condition expressed by Eq. (4). A similar procedure applied to the cylindrical configuration leads to the corresponding condition expressed by Eq. (9).

## 5. Discussion and conclusions

We have suggested in this work to exploit dielectric-metal core-shell nanostructures for efficient resonant and yet broadband absorption of infrared radiation, demonstrating highly efficient absorbers based on nm-thin Ti shells in spherical and cylindrical configurations. It is important to note that many other metals can also be found suitable for the same purpose because of similar dielectric properties. Thus, core-shell spherical and cylindrical configurations with 5-nm-thin shells made of Cr [25] and Ni [26] exhibit the absorption and scattering characteristics that are rather similar to those of Ti-based configurations (Fig. 4). The most noticeable difference is in the blue shift of the absorption and scattering maxima that can directly be deduced from the corresponding maximum conditions [cf. Eqs. (4) and (9)] when taking into account relatively high values of the dielectric constants of these materials. One should, however, bear in mind the possibility of encountering technological problems when trying to realize the corresponding efficient absorbers for longer wavelengths since that would require using even thinner metal shells. The important circumstance, that the optimal shell thickness scales with the dielectric constant of the core material [cf. Eqs. (4) and (9)], suggests, as a possible solution to that problem, the usage of high-index semiconductor or composite materials [27–29] as a core material.

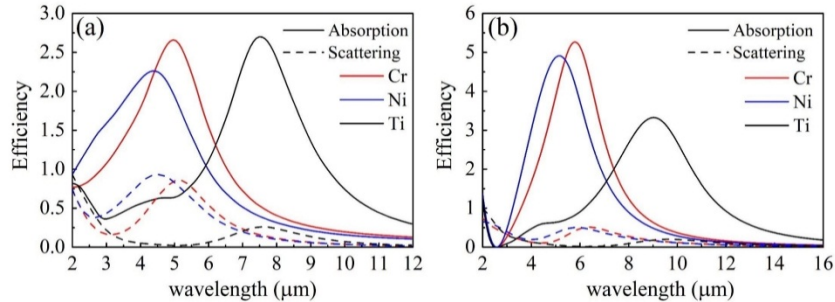

Fig. 4. Absorption (solid) and scattering (dashed) efficiency spectra of the (a) spherical and (b) cylindrical metal-shell configurations based on 5-nm-thin shells made of different metals: Ni (blue), Ti (black) and Cr (red). The core radius of 500 nm and permittivity $\varepsilon_1 = 2.25$ is used in all configurations.

Summarizing, we have suggested and investigated the usage of dielectric-metal core-shell nanostructures with nm-thin shells made of poor metals, i.e., metals having real and imaginary parts of their dielectric permittivities of the same order, for designing very efficient and broadband absorbers of infrared radiation. We have demonstrated that these structures can serve as very efficient and universal basic elements for developing broadband absorbers in the infrared, which would also ensure very low scattering, negligibly small in comparison with the absorption. It has been shown that the absorption characteristics can be controlled with the appropriate choice of the geometrical and material parameters used. . We have obtained simple analytical expressions for the absorption resonances in spherical and cylindrical configurations that allow one to quickly identify the configuration parameters ensuring strong infrared absorption in a given spectral range. Relations to effective medium parameters obtained by the internal homogenization are established and discussed. Finally, we would like to emphasize that the proposed nanostructures exhibit high degrees of symmetry, ensuring thereby that the absorption is completely (spherical) or partially (cylinder) insensitive to the angle and polarization of incident radiation. Unlike the case of traditional metamaterial absorbers in the infrared, there is no need for an orderly arrangement of these elements, an important feature that opens the prospect for practical use of these structures in random surface arrangements. Overall, we believe that our results provide new important physical insights into the design strategy for core-shell nanostructures and can be used as practical guidelines for realization of isotropic, efficient and broadband infrared absorbers of subwavelength sizes desirable in diverse applications, for example, in thermal detection and thermophotovoltaics.

## Funding

European Research Council (Advanced Grant PLAQNAP); University of Southern Denmark (SDU 2020); Villum Kann Rasmussen Foundation (Award in Technical and Natural Sciences 2019).